\documentclass[prl,aps,twocolumn,showpacs,floatfix]{revtex4-1}
\usepackage{amsfonts}
\usepackage{amssymb}
\usepackage{hyperref}
\usepackage{graphicx}
\usepackage{dcolumn}
\usepackage{bm,amsmath,verbatim}
\usepackage{mathrsfs}
\usepackage{color}
\usepackage{braket}

\usepackage{amsmath,amssymb,amsthm}

\usepackage[T1]{fontenc}
\usepackage[latin9]{inputenc}
\usepackage{booktabs}

\providecommand{\tabularnewline}{\\}
\renewcommand\toprule{\hline\hline}
\renewcommand\bottomrule{\hline\hline}

\begin{document}
\title{Exceptional Heavy-Fermion Semimetals in Three Dimensions}
\author{Yu-Liang Tao$^{1}$}
\author{Tao Qin$^{2}$}
\email{taoqin@ahu.edu.cn}
\author{Yong Xu$^{1,3}$}
\email{yongxuphy@tsinghua.edu.cn}

\affiliation{$^{1}$Center for Quantum Information, IIIS, Tsinghua University, Beijing 100084, People's Republic of China}
\affiliation{$^{2}$Department of Physics, School of Physics and Optoelectronics Engineering,
Anhui University, Hefei, Anhui Province 230601, People's Republic of China}
\affiliation{$^{3}$Shanghai Qi Zhi Institute, Shanghai 200232, People's Republic of China}

\begin{abstract}
Topological heavy-fermion systems in three dimensions are usually classified as topological insulators or semimetals.
Here, we theoretically predict a different type of heavy-fermion system (dubbed exceptional heavy-fermion semimetal)
by studying a three-dimensional periodic Anderson model consisting of strongly correlated localized $f$ electrons
and itinerant conduction $c$ electrons in a zincblende lattice. Due to the breaking of inversion symmetry,
the quasiparticle lifetimes at different sublattices are distinct, leading to the emergence of Weyl exceptional
rings in the complex pole of the Green's function at finite temperatures; such rings lead to the appearance
of bounded Fermi surfaces (bulk Fermi disks). As temperatures rise, two pairs of Weyl exceptional rings merge
into two exceptional rings with one bounded bulk Fermi surface (bulk Fermi tube), which are experimentally
measurable by angle-resolved photoemission spectroscopy. Finally, we use the dynamical mean field theory
to calculate the spectral functions which illustrate the emergence of bulk Fermi tubes.
Our work thus opens the door for studying exceptional heavy-fermion semimetal phases in three dimensions.
\end{abstract}

\maketitle

\section{Introduction}

Strongly correlated systems host a variety of intriguing phenomena beyond noninteracting electrons~\cite{ImadaRMP,DemlerRMP,PLeeRMP,TosattiRMP,SiNRP}.
For instance, strongly correlated systems may allow for the existence of a bulk Fermi arc with ending points,
which has been experimentally observed in the pseudogap phase of two-dimensional (2D) copper oxide high temperature superconductors~\cite{Keimer2015Nat}. Such a bulk Fermi surface with boundaries is not allowed in a Hermitian noninteracting
system with translational symmetry. Even in type-II Weyl semimetals, while a bulk Fermi surface can become open, boundaries are not allowed~\cite{Xu2015PRL,Soluyanov2015nat}. In this context, it has been theoretically shown that bulk Fermi arcs can also appear
in 2D heavy-fermion systems due to the presence of exceptional points~\cite{Yoshida2018PRB,Nagai2020PRL,Yoshida2019PRB,Kimura2019PRB,Michishita2020PRB,Michishita2020PRL}, where the single-particle
effective Hamiltonian becomes nondiagonalizable. However, it is not clear whether bulk Fermi surfaces with boundaries
can also emerge in a realistic three-dimensional (3D) strongly correlated material.
The question is motivated by
recent discovery of exceptional rings with bounded Fermi surfaces
in noninteracting non-Hermitian ultracold atomic systems or optical systems~\cite{Xu2017PRL,Cerjan2019nat,Cerjan2018PRB,Zyuzin2018PRB,Carlstrom2018PRA,HuPRB2019,Wang2019PRB,Kawabata2019PRL,Liu2021arxiv}.

\begin{figure}[t]
	\includegraphics[width = 1\linewidth]{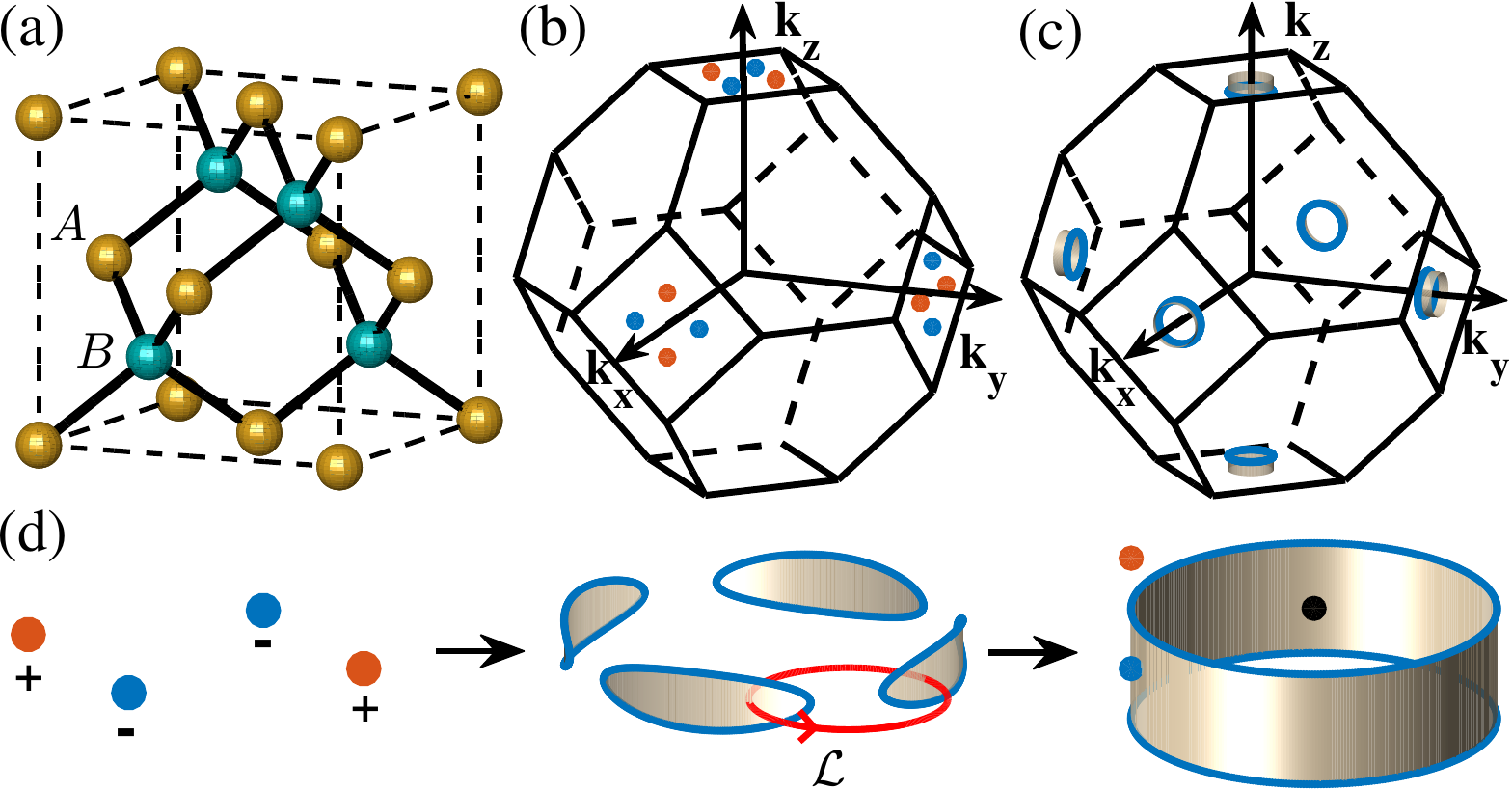}
	\caption{(Color online) (a) Schematic of the zincblende structure
consisting of $A$ and $B$ sublattices.
The first Brillouin zone of the fcc structure (b) with six pairs of Weyl points marked out as red (chiral charge $+1$)
and blue (chiral charge $-1$) solid circles, which develop into three bulk Fermi tubes as shown in (c).
(d) Schematic of the evolution of the zero-energy structure from four Weyl points to four Weyl exceptional rings
marked by blue curves
with bulk Fermi disks highlighted by the gold color;
the Weyl exceptional rings finally develop into two exceptional rings with bulk Fermi tubes highlighted by the gold color.
The winding number over the closed red circle enclosing two Weyl exceptional rings vanishes so that the two rings can merge.}	
	\label{Fig1}
\end{figure}

In heavy-fermion materials, apart from topological Kondo insulating phases~\cite{Coleman2010PRL,Galitski2012PRB,Dai2013PRL},
Weyl semimetal phases may also emerge,
such as in noncentrosymmetric CeRu$_4$Sn$_6$ or Ce$_3$Bi$_4$Pd$_3$~\cite{Dai2017PRX,Lai2018PNAS,Grefe2020PRB,Dzsaber2017PRL,Dzsaber2021PNAS,
Steglich2010PRB,Guritanu2013PRB,Sundermann2015SR,Held2016EPJB}.
In the paper,
we study a microscopic periodic Anderson model (PAM) describing an $f$ electron
system, such as CeRu$_4$Sn$_6$ or Ce$_3$Bi$_4$Pd$_3$,
and theoretically predict a different type of heavy-fermion state: exceptional heavy-fermion
semimetals which have exceptional rings with bounded Fermi surfaces
in the complex pole of the Green's function
at finite temperatures.
The model consists of strongly correlated localized $f$ electrons and itinerant conduction electrons in a zincblende structure
with $A$ and $B$ sublattices [see Fig.~\ref{Fig1}(a)]. The interactions for $f$ electrons renormalize the effective one-body Hamiltonian through
a self-energy in the retarded Green's function. In the presence of hybridization between $f$ electrons and conduction
electrons, the interactions not only renormalize parameters for a Weyl Hamiltonian but also transform Weyl points
into exceptional rings. Such a ring arises from the fact that $f$ electrons on $A$ and $B$ sublattices
exhibit different lifetimes due to the broken inversion symmetry.
Based on the second-order perturbation theory, we show that a Weyl point develops into a Weyl exceptional ring with a bulk Fermi disk as temperatures rise [see Fig.~\ref{Fig1}(d)].
As we further raise temperatures, two pairs of such Weyl exceptional rings merge into two exceptional rings [see Fig.~\ref{Fig1}(d)],
leading to the emergence of a bounded Fermi surface in the shape of a tube. Finally, we utilize the dynamical mean-field theory (DMFT) to numerically
compute the spectral functions illustrating the emergence of the bulk Fermi tubes in our system.
Given the fact that a noncentrosymmetric heavy fermion semimetal Ce$_3$Bi$_4$Pd$_3$ has been
experimentally identified~\cite{Dzsaber2017PRL,Dzsaber2021PNAS}, we expect that the Fermi tubes may be experimentally observed in the material.

\section{Periodic Anderson models}

We start by considering a 3D periodic Anderson model consisting of strongly correlated
localized $f$ electrons and conduction $c$ electrons in a zincblende lattice with two sublattices denoted by $A$ and $B$.
The Hamiltonian reads
\begin{equation} \label{H_all}
\hat{H}=\hat{H}_c+\hat{H}_f+\hat{H}_{cf},
\end{equation}
where $\hat{H}_c$, $\hat{H}_f$ and $\hat{H}_{cf}$ describe the conduction $c$ electrons, localized $f$ electrons and their hybridization, respectively.
Specifically, $\hat{H}_f=\varepsilon_f\sum_{j,\sigma}\hat{f}^{\dagger}_{j\sigma}\hat{f}_{j\sigma}+U\sum_j \hat{n}^f_{j\uparrow}\hat{n}^f_{j\downarrow}$
with $\varepsilon_f$ being the energy of localized $f$ electrons and $U$ characterizing the Coulomb repulsion strength
for $f$ electrons, and $\hat{H}_{cf}=V\sum_{j,\sigma}(\hat{f}^{\dagger}_{j\sigma}\hat{c}_{j\sigma}+{H.c.})$ with $V$ denoting the
hybridization strength. Here, $\hat{c}_{j\sigma}$ and $\hat{f}_{j\sigma}$ [$\hat{c}_{j\sigma}^\dagger$ and $\hat{f}_{j\sigma}^\dagger$] are the
fermion annihilation (creation) operators for a conduction and
$f$ electron with spin $\sigma$ at site $j$, respectively;
$\hat{n}^f_{j\sigma}$ refers to the number of $f$ electrons with spin $\sigma$ at site $j$.
For conduction electrons,
we write down its Hamiltonian in momentum space as
$\hat{H}_c=\sum_{\bm k}\hat{\Psi} ^{\dagger}_{\bm k} H_c({\bm k}){\hat{\Psi}_{\bm k}}$,
where
$\hat{\Psi}^\dagger_{\bm k}=(\begin{array}{cccc}
                  \hat{c}^\dagger_{{\bm k}\uparrow,A} & \hat{c}^\dagger_{{\bm k}\uparrow,B} & \hat{c}^\dagger_{{\bm k}\downarrow,A} & \hat{c}^\dagger_{{\bm k}\downarrow,B}
                \end{array})$ and
\begin{equation}
\label{H_c}
H_c({\bm k})=\sigma_0[u_1({\bm k})\tau_x+u_2({\bm k})\tau_y+m\tau_z]+
\lambda{\bm D}\cdot{\bm \sigma}\tau_z,
\end{equation}
which is a modified Fu-Kane-Mele model~\cite{Fu2007PRL}. Here, $\sigma_\nu$ and $\tau_\nu$ ($\nu=x,y,z$) represent Pauli matrices
acting on spin and sublattice degrees of freedom, respectively, and $\sigma_0$ is the identity matrix.
$u_1({\bm k})$ and $u_2({\bm k})$ are determined by the nearest-neighbor hopping between different sublattices with strength $t$,
and $2m$ represents the amount of the on-site potential difference on sublattices $A$ and $B$, which breaks inversion symmetry. $D_\nu({\bm k})$ ($\nu=x,y,z$)
is determined by the spin-orbit coupling with strength $\lambda$. The specific expressions
of $u_1$, $u_2$ and $D_\nu$ can be found in Appendix A. To simplify notations, we have set the lattice constant $a=1$.

\begin{figure}[t]
	\includegraphics[width = 1\linewidth]{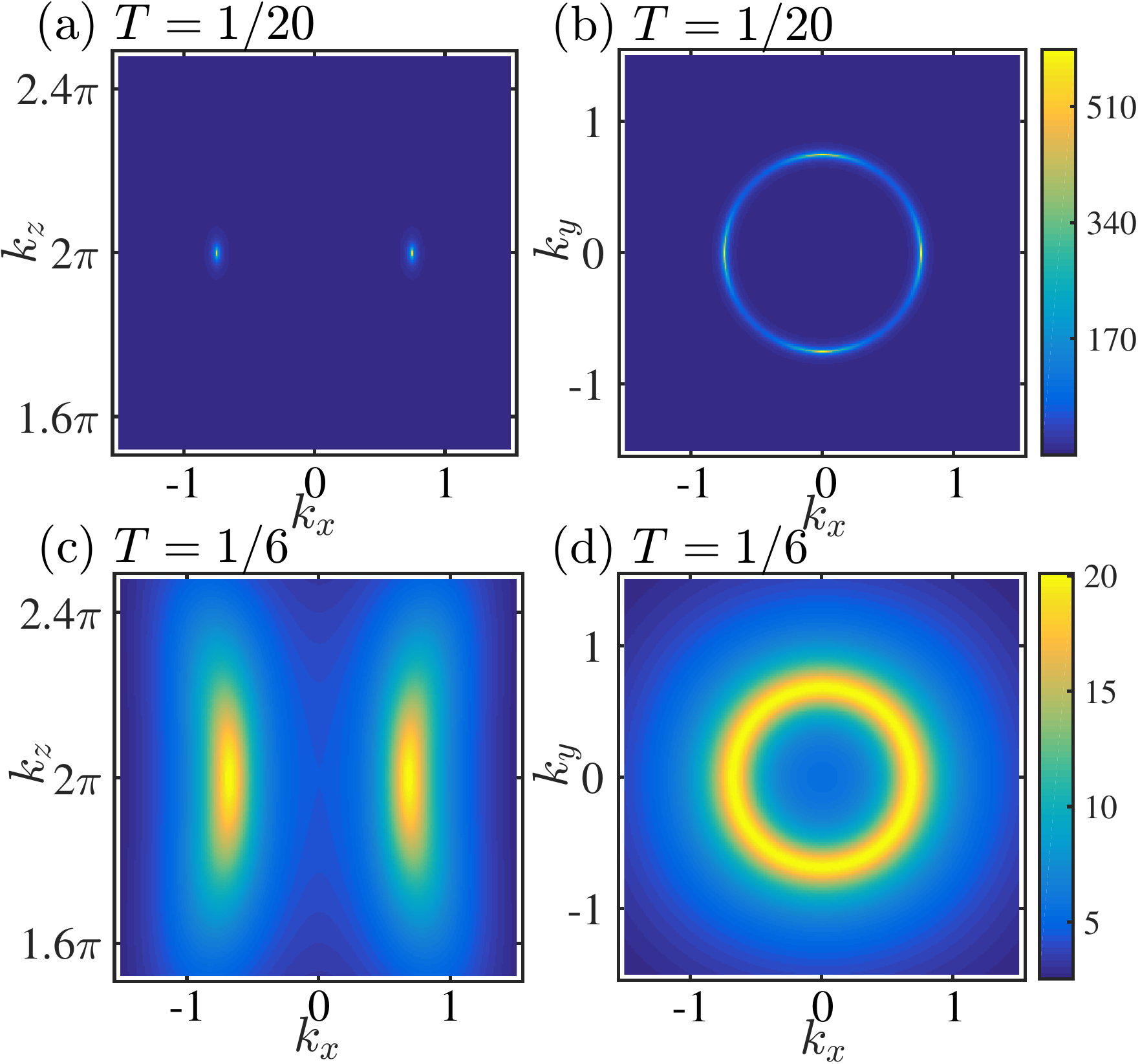}
	\caption{(Color online) {
		(a) The first Brillouin zone of the fcc structure. The red lines denote the high symmetry path. (b) The band structure of the Hamiltonian
		$H_c ({\bm k})$ of conduction c electrons along the high symmetry path. The band structure of the whole Hamiltonian $H_0 ({\bm k})$ along the high symmetry path
		(c) without $\varepsilon_{s}$, and (d) with
		$\varepsilon_{s}$.
		Here, $t=0.5$, $m=1.2$, $\lambda=0.6$, $V=2$, $\varepsilon_f=1$ and $\varepsilon_{s}=4$.   }
	}
	\label{Fig2-1}
\end{figure}

Without interactions, when $|m|<4|\lambda|$, the Hamiltonian of conduction electrons exhibit six pairs of Weyl points located at $(\pm k_0,0,2\pi)$, $(2\pi,0,\pm k_0)$, $(0,\pm k_0,2\pi)$, $(0,2\pi,\pm k_0)$, $(\pm k_0,2\pi,0)$ and $(2\pi,\pm k_0,0)$,
where ${k_0}={2}\sin^{-1}(|{m}/{4\lambda}|)$ with $0<k_0<\pi$ [see the locations of Weyl points in the first Brillouin zone in Fig.~\ref{Fig1}(b)
and the band structure in Fig.~\ref{Fig2-1}(b)].
These points annihilate each other through the critical point $|m|=4|\lambda|$, leading to a topologically trivial insulator when
$|m|>4|\lambda|$.

In the presence of localized $f$ electrons and the hybridization between $f$ and $c$ electrons, the Hamiltonian in momentum space without interactions is expressed as
\begin{equation}
\label{H_all_simple}
H_0({\bm k})=
\left(
\begin{array}{cc}
\varepsilon_f  & {V }\\
{V} &H_c({\bm k})
\end{array}
\right).
\end{equation}
The hybridization changes the energy $\varepsilon_{c,i}({\bm k})$ ($i=1,2,3,4$) of $H_c({\bm k})$ to two energies
$\varepsilon_{i,\pm}({\bm k})=[(\varepsilon_f+\varepsilon_{c,i}({\bm k}))\pm \sqrt{(\varepsilon_f-\varepsilon_{c,i}({\bm k}))^2+4V^2}]/2$.
Thus, a Weyl point at zero energy in $H_c({\bm k})$ becomes two Weyl points with different energies: One has
a negative energy corresponding to a quarter filling [see Fig.~\ref{Fig2-1}(c)]. For convenience, we will add a constant energy shift
$\varepsilon_{s}=V^2/\varepsilon_f$ in $H_c$ so that the energy at Weyl points between the second and third bands
is fixed at the zero energy [see Fig.~\ref{Fig2-1}(d)]. Note that such a shift will not change the physics.

\begin{figure}[t]
	\includegraphics[width = 1\linewidth]{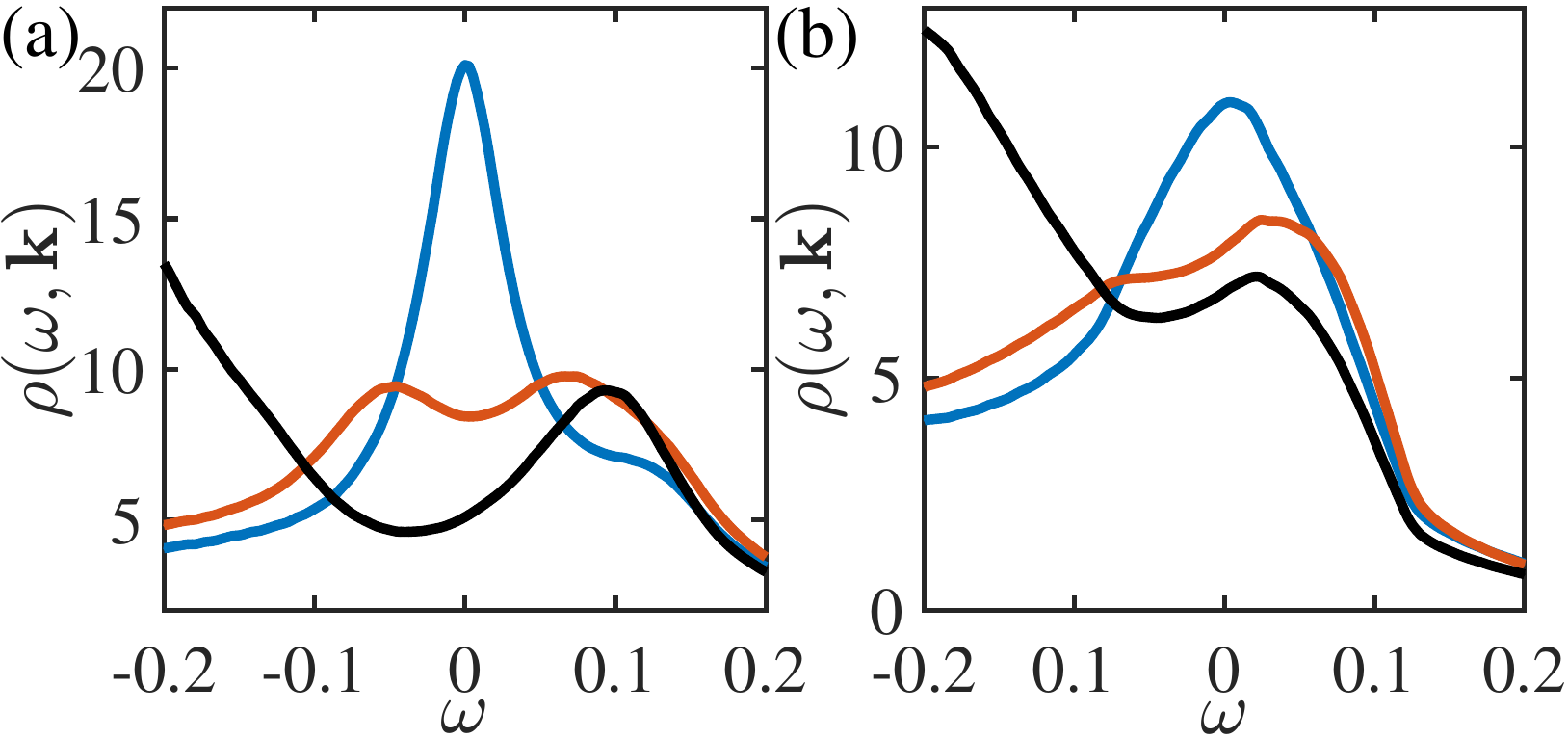}
	\caption{(Color online)
		The sectional view of the zero-energy spectral functions (a)(c) in the $k_y=0$ plane around $k_z=2\pi$ and (b)(d) in the $k_z=2\pi$ plane, which are calculated by the perturbation theory.
		The results imply the existence of bulk Fermi disks in (a-b) or bulk Fermi tubes in (c-d) due to the appearance of Weyl exceptional rings
		[see Fig.~\ref{Fig3-1}(a)] at the temperature $T=1/20$
		or a pair of exceptional rings [see Fig.~\ref{Fig3-1}(b)] at $T=1/6$, respectively.
		Here, $t=0.5$, $m=1.2$, $\lambda=0.6$, $V=2$, $U=2$, $\varepsilon_f=1$ and $\varepsilon_{s}=4$.
	}
	\label{Fig2}
\end{figure}

In the presence of interactions, we consider the retarded Green's function at the energy $\omega$
\begin{equation} \label{Sigma_eq}
G^R(\omega,{\bm k})=[\omega+\mu-H_0({\bm k})-\Sigma(\omega,{\bm k})]^{-1},
\end{equation}
where $\mu$ is the chemical potential and $\Sigma(\omega,{\bm k})$ is the self-energy. Similar to
the two-dimensional case~\cite{Nagai2020PRL}, since there are interactions only for $f$ electrons, only $f$ electrons
acquire a nonzero self-energy,
\begin{equation}
\Sigma(\omega)=\left(
                         \begin{array}{cc}
                           \Sigma^f(\omega) & 0 \\
                           0 & 0 \\
                         \end{array}
                       \right).
\end{equation}
Here, we also assume that the self-energy is independent of quasimomenta because we consider heavy $f$ electrons without
dispersion (in other words, the temperature is high compared to the bandwidth of $f$ electrons)~\cite{Nagai2020PRL}.
With time-reversal symmetry, $\Sigma^f$ is independent of spins, i.e.,
$[\Sigma^f]_{\sigma\sigma^\prime}=[\Sigma^f]_{\sigma\sigma^\prime}\delta_{\sigma \sigma^\prime}$.
However, without inversion symmetry, $\Sigma^f$ can have different components at different sublattices.
At finite temperatures, the self-energy takes complex values due to quasiparticle finite lifetimes. The breaking of inversion
symmetry thus leads to different lifetimes for electrons at different sublattices, resulting
in the appearance of Weyl exceptional rings as shown in the following discussion.

To demonstrate that exceptional rings emerge in the presence of lifetime difference of electrons
at different sublattices, we expand the self-energy in the Taylor series up to the first order
with respect to $\omega$,
\begin{equation}
\Sigma^f(\omega) \approx a_{0}-i\Gamma_{0}+(a_{1}-i\Gamma_{1})\tau_{z}+a_{0\omega}\omega+a_{1\omega}\omega\tau_{z},
\end{equation}
where $a_0+a_1$ and $a_0-a_1$ ($a_{0\omega}+a_{1\omega}$ and $a_{0\omega}-a_{1\omega}$) describe the zeroth-order
(first-order) real parts of the self-energy at sublattices $A$ and $B$, respectively,
and $\Gamma_0+\Gamma_1$ and $\Gamma_0-\Gamma_1$ depict the inverse of quasiparticle lifetimes at sublattices $A$
and $B$, respectively. To present clearly, we here do not consider the imaginary contribution in the first-order
correction (see Appendix B for the derivation). In this case, the first-order terms only renormalize parameters as
$\varepsilon_{fr}=\varepsilon_f+a_0\rightarrow \bar{\varepsilon}_{fr}=[(Z_A+Z_B)\varepsilon_{fr}+(Z_A-Z_B)a_1]/2$,
$\Gamma_0\rightarrow \bar{\Gamma}_0=[(Z_A+Z_B)\Gamma_0+(Z_A-Z_B)\Gamma_1]/2$,
$a_1\rightarrow \bar{a}_1=[(Z_A-Z_B)\varepsilon_{fr}+(Z_A+Z_B)a_1]/2$, $\Gamma_1\rightarrow \bar{\Gamma}_1=[(Z_A-Z_B)\Gamma_0+(Z_A+Z_B)\Gamma_1]/2$ with $Z_A=1/(1-a_{0\omega}-a_{1\omega})$ and $Z_B=1/(1-a_{0\omega}+a_{1\omega})$. The first-order terms also renormalize
the coupling matrix $\text{diag}(V,V)$ to $\text{diag}(V_A,V_B)$ with $V_A=\sqrt{Z_A}V$ and $V_B=\sqrt{Z_B}V$.
When $\Gamma_0=\Gamma_1=0$, we add an energy shift $\varepsilon_s=(\bar{\varepsilon}_{fr}\bar{V}^2+V_0 \bar{a}_1)/(\bar{\varepsilon}_{fr}^2-\bar{a}_1^2)$ with $\bar{V}=\sqrt{V_1^2+V_2^2}$, $V_0=-2V_1 V_2$, $V_1=(V_A+V_B)/2$ and
$V_2=(V_A-V_B)/2$ in $H_c$ to fix the energy of Weyl points at zero
corresponding to a quarter filling. There, the locations of Weyl points in momentum space are still determined
by $H_c$ with a renormalized mass $\bar{m}=m-d_{z0}$ with $d_{z0}=(\bar{V}^2 a_1+V_0\bar{\varepsilon}_{fr})/(a_1^2-\bar{\varepsilon}_{fr}^2)$. In fact, only $k_0$
is changed to ${\bar{k}_0}={2}\sin^{-1}(|\bar{m}/{4\lambda}|)$ with $0<\bar{k}_0<\pi$.

\begin{figure}[t]
	\includegraphics[width = 1\linewidth]{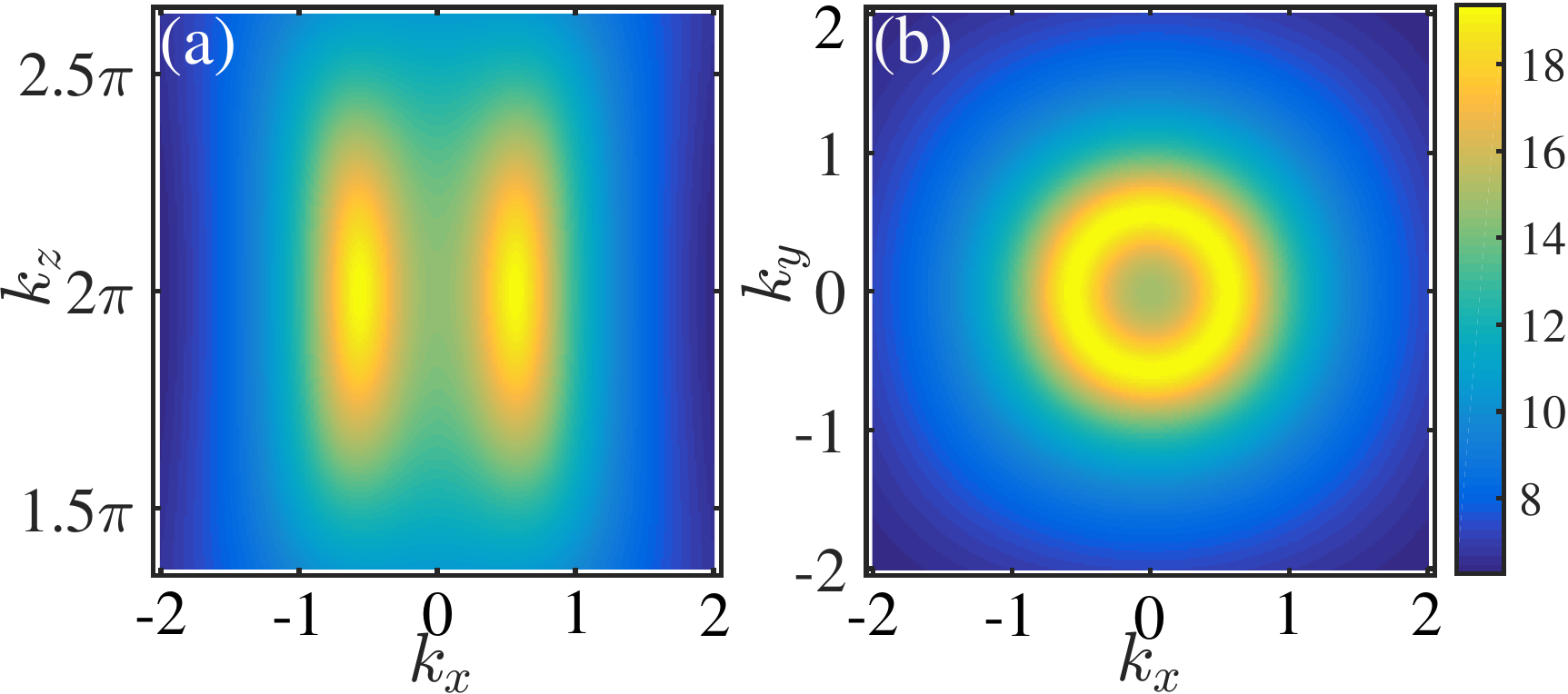}
	\caption{(Color online) The contours of exceptional rings around $k_z=2\pi$ at the temperature (a) $T=1/20$ and (b) $T=1/6$,
		which are calculated by the perturbation theory with an approximation, $a_{0 \omega}\approx 0$ and $a_{1 \omega}\approx 0$.
		Blue/red nodes denote the original Weyl points.
		Here, $t=0.5$, $m=1.2$, $\lambda=0.6$, $V=2$, $U=2$, $\varepsilon_f=1$ and $\varepsilon_{s}=4$.
	}
	\label{Fig3-1}
\end{figure}

We now study the effects of the imaginary parts of the self-energy on the pole of the Green's function. To derive an analytical expression of
the energy close to a Weyl point,
we assume that $\Gamma_0$, $\Gamma_1$, $a_1$, $a_{0\omega}$
and $a_{1\omega}$ are small quantities.
Slightly away from a Weyl point ${\bm k}_W$ determined by $u_1({\bm k}_W)=u_2({\bm k}_W)=0$ and $\bar{m}+\alpha \lambda D({\bm k}_W)=0$ with
$D=\sqrt{D_x^2+D_y^2+D_z^2}$ and
$\alpha=\pm 1$,
$u_1$, $u_2$ and $\bar{m}+\alpha \lambda D({\bm k}_W)$ are small quantities.
Specifically,
$u_1({\bm k}_W+\delta{\bm k})=d_x$, $u_2({\bm k}_W+\delta{\bm k})=d_y$
and $\bar{m}+\alpha \lambda D({\bm k}_W+\delta{\bm k})=d_z$,
where $\delta {\bm k}$ is a small vector
measured with respect to ${\bm k}_W$. The energy is derived as
\begin{equation}
\omega=-i\bar{\Gamma}_{0}\bar{v}_0
\pm\sqrt{\bar{v}_1^2[d_x^2+d_y^2+(d_{z}-i\bar{\gamma}_{0})^{2}]},
\end{equation}
where $\bar{v}_0={\varepsilon_{s}}/({\bar{\varepsilon}_{fr}+\varepsilon_{s}})$, $\bar{v}_1={\bar{\varepsilon}_{fr}}/({\bar{\varepsilon}_{fr}+\varepsilon_{s}})$
and $\bar{\gamma}_{0}={\varepsilon_{s}\bar{\Gamma}_{1}}/{\bar{\varepsilon}_{fr}}$. Remarkably, the inverse lifetime difference $\Gamma_1$
at two sublattices leads to the emergence of a Weyl exceptional ring determined by $d_z=0$ and $d_x^2+d_y^2=\gamma_0^2$, where
the Hamiltonian becomes nondiagonalizable. One of the authors has established that a Weyl exceptional ring is characterized by two topological invariants: the Chern number and the Berry phase~\cite{Xu2017PRL}.
In addition, the real part of the energy vanishes inside the ring, leading to a bulk Fermi surface
in the shape of a Fermi disk.
Specifically, consider the two pairs of Weyl points on the $k_z=2\pi$ plane.
Based on the perturbation theory up to the second order (see Appendix C for details),
as temperatures rise, the difference of the inverse of quasiparticle lifetimes $\Gamma_1$ gets bigger,
leading to the development of four Weyl exceptional rings from four Weyl points;
as $\Gamma_1$ further increases,
the neighboring Weyl exceptional rings merge and become two exceptional rings [see Fig.~\ref{Fig1}(d)].
The two rings serve as two boundaries of a bounded Fermi surface in the shape of a Fermi tube [there is a total of three Fermi tubes in the
first Brillouin zone as shown in Fig.~\ref{Fig1}(c)].
The mergence can happen due to the fact that the winding number defined as ~\cite{Shen2018PRL2,Ueda2018PRX,Zeng2020PRB}
\begin{align}
	\label{windingnumber}
	W_{\mathcal{L}}=\frac{1}{2\pi}\oint_{\mathcal{L}} d{\bm k} \cdot \nabla_{{\bm k}} [\text{arg}(\omega_+)+\text{arg}(\omega_-)],
\end{align}
vanishes over a closed path enclosing two neighboring rings [see Fig.~\ref{Fig1}(d)]. Here,
$\omega_-$ and $\omega_+$ refer to the two energies close to zero energy which are numerically obtained by approximating
the self-energy up to the first order.

\begin{figure}[t]
	\includegraphics[width = 0.94\linewidth]{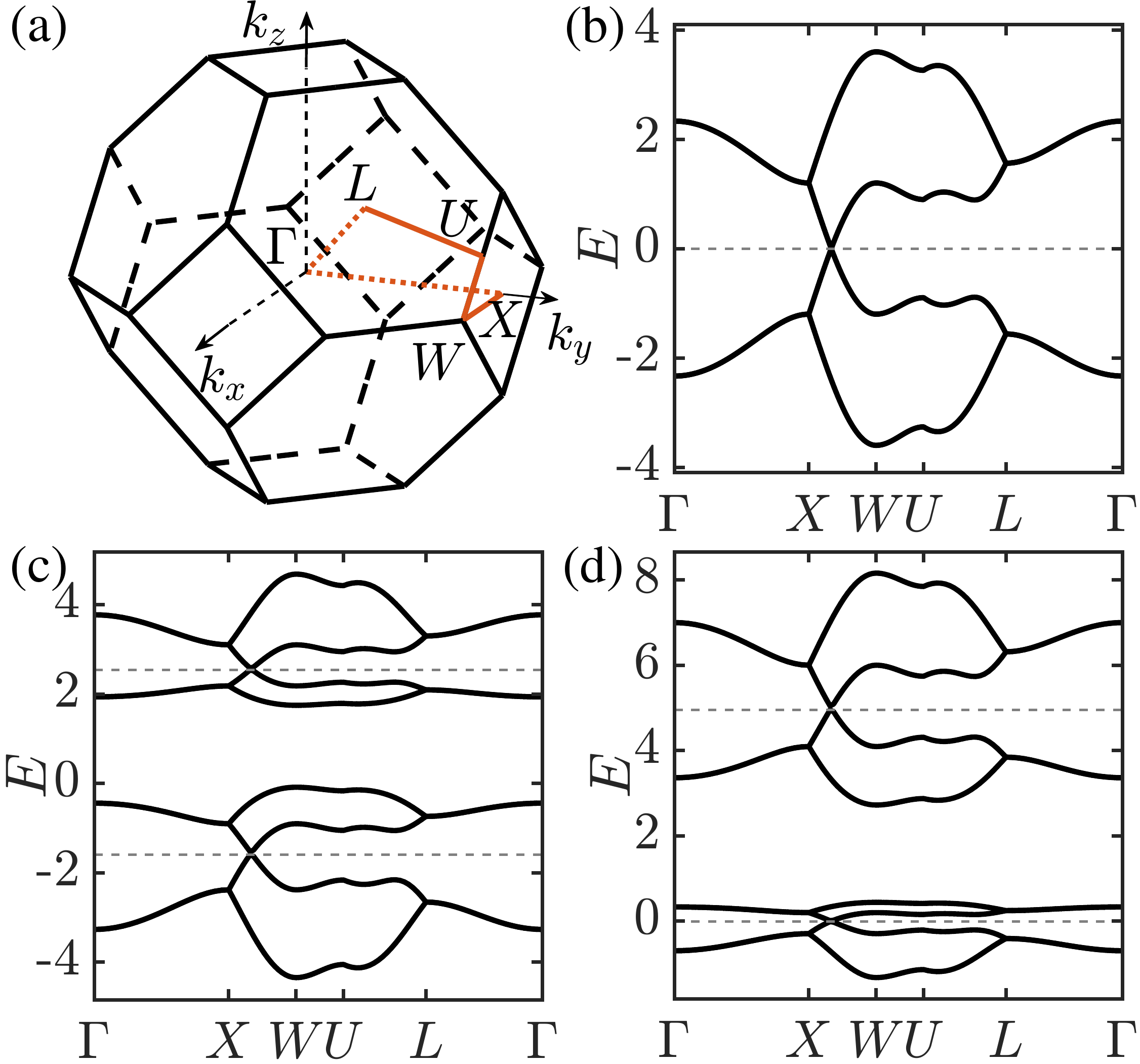}
	\caption{(Color online)
The spectral function versus the energy for three typical points in momentum space.
The positions of these points are schematically marked out by the corresponding colored solid circles in
Fig.~\ref{Fig1}(d) (right).
(a) is calculated by the perturbation theory with the same system parameters as in Fig.~\ref{Fig2} at $T=1/6$,
and (b) is calculated by the DMFT at $T=1/11$ with $t=0.5$, $m=1.2$, $\lambda=0.6$, $V=2$, $U=2.5$, $\varepsilon_f=0.275$ and $\varepsilon_{s}=4$.
	}
	\label{Fig3}
\end{figure}

The bounded Fermi surface manifests in the spectral functions, which can be experimentally measured by angle-resolved
photoemission spectroscopy (ARPES). The spectral functions read
\begin{equation}
\rho(\omega,\mathbf{k})=-(1/\pi) \textrm{ImTr}[G^R(\omega,\mathbf{k})],
\end{equation}
which reflects the pole information of the Green's function.
To demonstrate, we calculate the self-energy by the perturbation theory up to
the second order (see Supplemental Material for details) and then evaluate the spectral functions (see Fig.~\ref{Fig2}).
Specifically, when $T=1/20$, there are two pairs of Weyl exceptional rings with four bulk Fermi disks [see Fig.~\ref{Fig1}(d) (center) and {\color{red}Fig.~\ref{Fig3-1}(a)}].
Note that in Figs.~\ref{Fig2}--\ref{Fig4}, we take $k_B\cdot \text{Kelvin}$ and Kelvin as energy and temperature units, respectively.
We illustrate the Fermi disk structures by the sectional view of the zero-energy spectral function
in the $k_y=0$ plane around $k_z=2\pi$ and in the $k_z=2\pi$ plane. The former shows
two short bright lines and the latter shows four bright arcs.
The arcs are connected to form a ring with much smaller values in the connecting parts,
which arises from the fact that the existence of $\Gamma_0$ widens the spectral functions.
When the temperature is raised to $T=1/6$, two pairs of Weyl exceptional rings become two rings with a bulk Fermi tube [{\color{red}Fig.~\ref{Fig3-1}(b)}].
Similarly, the sectional view of the zero-energy spectral function reflects the bulk Fermi tube structure:
There are two bright lines in the $k_y=0$ plane and a bright circle in the $k_z=2\pi$ plane.
Figure~\ref{Fig3}(a) further displays the spectral functions with respect to the energy for three typical points
in momentum space. On the Fermi tube, the spectral function exhibits a peak at zero energy;
away from the tube, it develops a minimum around the zero energy and peaks away from the zero energy,
consistent with the energy spectrum structure (see Appendix D for details).

\section{Spectral functions calculated by the DMFT}

In order to analyze the interacting effects more accurately, we adopt the DMFT
with the segment-based hybridization-expansion continuous-time quantum Monte Carlo impurity solver (CT-HYB) implemented in the toolkit Triqs~\cite{Parcollet2015CPC}. Within the DMFT, we treat the self-energy $\Sigma(\bm{k},\omega)$ in Eq.(\ref{Sigma_eq}) approximately as $\Sigma(\omega)$ based on the local fluctuation approximation. We also numerically confirm that the off-diagonal entries in $\Sigma^f(\omega)$ are much smaller than the diagonal ones.
Even though the self-energy is $\bm{k}$-independent, it is beyond the reach of the perturbation theory for intermediate and strong interactions.

To calculate the spectral function $\rho(\omega,\bm{k})$, we first employ the DMFT to compute the imaginary time Green's function
and then carry out the numerical analytic continuation of the imaginary time self-energy
$\Sigma(i\omega_n)$ with Triqs/maxent.
For the numerical analytic continuation, we find that the output of $\Sigma(\omega)$ is extremely sensitive to the noise in $\Sigma(i\omega_n)$. To ensure the reliability of our results, we need to reduce the amplitude of noises as far as possible. In practice, we utilize the Legendre polynomial to reduce high-frequency noises during self-consistent iterations and average multi-step iterative results of $\Sigma(i\omega_n)$ as the final output after convergence.

\begin{figure}[t]
	\includegraphics[width = 1\linewidth]{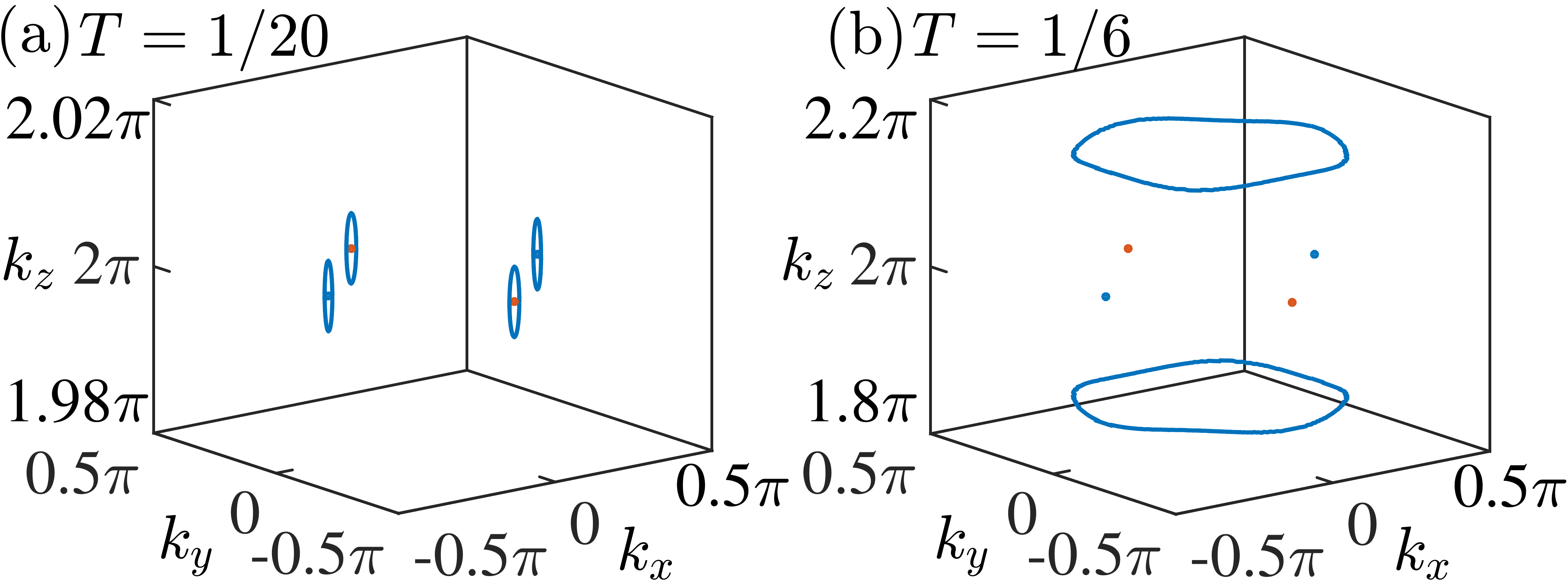}
	\caption{(Color online)
The sectional view of the zero-energy spectral functions (a) in the $k_y=0$ plane around $k_z=2\pi$ and (b) in the $k_z=2\pi$ plane,
which are calculated by the DMFT at $T=1/11$.
The results indicate the existence of bulk Fermi tubes due to the appearance of a pair of exceptional rings.
Here, $t=0.5$, $m=1.2$, $\lambda=0.6$, $V=2$, $U=2.5$, $\varepsilon_f=0.275$ and $\varepsilon_{s}=4$.
	}
	\label{Fig4}
\end{figure}

Figure~\ref{Fig4} demonstrates the sectional view of the zero-energy spectral functions around $k_z=2\pi$ at $T=1/11$
obtained by the DMFT calculation. We see clearly the existence of bulk Fermi tubes, which is consistent with
the results computed by the perturbation theory.
The spectral functions with respect to the energy exhibit a peak at zero energy in a momentum on a
Fermi tube and peaks away from zero energy in momenta away from the Fermi tube [see Fig.~\ref{Fig3}(b)].
The results are qualitatively consistent with those obtained by the perturbation theory.
However, compared with the results from the perturbation theory calculations, the contrast
between the values of the peak and background of $\rho(\omega,\bm{k})$ from the DMFT calculations is lower.
We attribute this to a smaller value of $\Gamma_1/\Gamma_0$ from the DMFT than that from the perturbation theory,
generating a relatively larger background.
Because the DMFT is a better way to treat interactions, one can attribute the features above to interaction effects.
Clearly, the DMFT reveals that the bulk Fermi tubes benefit from the complex-valued self-energy from intermediate interactions.
However, one can expect that this interesting phenomenon would be finally suppressed by strong interactions when the system enters
into the Mott insulator phase (see Appendix E for details).

\section{Conclusion}

In summary, we have found a different type of 3D heavy-fermion phase: exceptional heavy-fermion semimetals which possess exceptional rings in the complex pole of the Green's function at finite temperatures.
Such rings give rise to bounded bulk Fermi surfaces such as Fermi disks or Fermi tubes manifesting in the spectral functions, which are experimentally measurable by ARPES. We finally use the dynamical mean field theory to calculate the spectral functions in our system,
revealing the emergence of bulk Fermi tubes. Recently, a noncentrosymmetric heavy fermion semimetal Ce$_3$Bi$_4$Pd$_3$ has been
experimentally identified~\cite{Dzsaber2017PRL,Dzsaber2021PNAS}, and we may expect that bulk Fermi tubes may be experimentally observed in the material.
Our work thus opens a new direction for studying exceptional heavy-fermion semimetal phases in 3D.

\begin{acknowledgments}
We thank Adriano Amaricci, Liang Du, Michel Ferrero, Li Huang, Kai Li, Yuki Nagai, Yilin Wang, Nils Wentzell, and Yan-Bin Yang for helpful discussions and communications.
This work is supported by the National Natural Science Foundation of China (Grants No. 11974201, U2032164) and Tsinghua University Dushi Program..
\end{acknowledgments}

\section*{Appendix A: Hopping terms in the Hamiltonian} 
\setcounter{equation}{0} 
\setcounter{table}{0}
\renewcommand{\theequation}{A\arabic{equation}} 
\renewcommand{\bibnumfmt}[1]{[#1]} \renewcommand{\citenumfont}[1]{#1}

The hopping terms $u_1$, $u_2$ and $D_\nu$ ($\nu=x,y,z$) in the Hamiltonian (2) in the main text are given by
\begin{widetext}
\begin{align}
	\label{hoppingterm}
	u_1({\bm k})=&t[1+\sum_{n=1}^{3}\cos({\bm k}\cdot{\bm a}_n)] \\
	u_2({\bm k})=&t\sum_{n=1}^{3}\sin({\bm k}\cdot{\bm a}_n) \\
	D_x({\bm k})=&\lambda\left\{\sin({\bm k}\cdot{\bm a}_2)-\sin({\bm k}\cdot{\bm a}_3)
	\sin\left[ {\bm k}\cdot({\bm a}_2-{\bm a}_1) \right] +\sin\left[ {\bm k}\cdot({\bm a}_3-{\bm a}_1) \right] \right\} \\
	D_y({\bm k})=&\lambda\left\{\sin({\bm k}\cdot{\bm a}_3)-\sin({\bm k}\cdot{\bm a}_1)
	\sin\left[ {\bm k}\cdot({\bm a}_3-{\bm a}_2) \right] +\sin\left[ {\bm k}\cdot({\bm a}_1-{\bm a}_2) \right] \right\} \\
	D_z({\bm k})=&\lambda\left\{\sin({\bm k}\cdot{\bm a}_1)-\sin({\bm k}\cdot{\bm a}_2)
	\sin\left[ {\bm k}\cdot({\bm a}_1-{\bm a}_3) \right] +\sin\left[ {\bm k}\cdot({\bm a}_2-{\bm a}_3) \right] \right\},
\end{align}
\end{widetext}
where ${\bm a}_1=(0,1/2,1/2)$, ${\bm a}_2=(1/2,0,1/2)$ and ${\bm a}_3=(1/2,1/2,0)$ are the lattice vectors for a fcc lattice.

\section*{Appendix B: The energy dispersion in the presence of the self-energy}
\setcounter{equation}{0} 
\setcounter{table}{0}
\renewcommand{\theequation}{B\arabic{equation}} 
\renewcommand{\bibnumfmt}[1]{[#1]} \renewcommand{\citenumfont}[1]{#1}

In this appendix, we will derive the energy dispersion near a Weyl point in the presence of the self-energy. For clarity, we first
study a simple case where the self-energy contains only the terms that are independent of the energy,
and show the emergence of Weyl exceptional rings arising
from the quasiparticle lifetime difference at different sublattices. After that, we demonstrate that the effects of
including a term in the self-energy that is linearly dependent of the energy is the
renormalization of system parameters, which does not affect the qualitative feature of the energy spectrum.

\subsection{B1. Energy spectra in the presence of the real energy independent self-energy}
We now study the effects of the terms in the self-energy that are independent of the energy, which read
\begin{equation}
	\Sigma^{f}=a_{0}-i\Gamma_{0}+(a_{1}-i\Gamma_{1})\tau_{z}\sigma_{0},
\end{equation}
where $a_0+a_1$ and $a_0-a_1$ denote the zeroth-order real parts of the self-energy at sublattices $A$ and $B$, respectively,
and $\Gamma_0+\Gamma_1$ and $\Gamma_0-\Gamma_1$ denote the inverse of quasiparticle lifetimes at sublattices $A$
and $B$, respectively. In the derivation, we first consider the complex self-energy and then make $\Gamma_0$ and $\Gamma_1$ zero.
The inverse of the Green's function is
\begin{equation}
	G^{-1}=\omega-\left(\begin{array}{cc}
		\tilde{\varepsilon}_{f}+a\tau_{z} & V\\
		V & H_{c}+\varepsilon_{s}
	\end{array}\right),
\end{equation}
where $\tilde{\varepsilon}_{f}=\varepsilon_{f}+a_{0}-i\Gamma_{0}=\varepsilon_{fr}-i\Gamma_{0}$
and $a=a_{1}-i\Gamma_{1}$. Here
\begin{equation}
	H_{c}=\sigma_{0}(u_{1}\tau_{x}+u_{2}\tau_{y}+m\tau_{z})+\lambda\bm{D}\cdot\bm{\sigma}\tau_{z}.
\end{equation}
We can transform this matrix into a block form
\begin{equation}
	\tilde{H}_{c}=S^{\dagger}H_{c}S=\left(\begin{array}{cc}
		h_{+} & 0\\
		0 & h_{-}
	\end{array}\right)=\bm{u}\cdot\bm{\tau}+\lambda D\sigma_{z}\tau_{z}
\end{equation}
with
$
h_{\pm}=u_{1}\tau_{x}+u_{2}\tau_{y}+(m\pm D\lambda)\tau_{z}
$,
$u_{x}=u_{1}$, $u_{y}=u_{2}$, and $u_{z}=m$ by the matrix
\begin{equation}
	S=\left(\begin{array}{cc}
		|u_{+}\rangle & |u_{-}\rangle\end{array}\right)\tau_{0}.
\end{equation}
Here
$|u_{\pm}\rangle$ are eigenstates of $\bm{D}\cdot\bm{\sigma}$
corresponding to eigenvalues $\pm D$, i.e.,
$\bm{D}\cdot\bm{\sigma}|u_{\pm}\rangle=\pm D|u_{\pm}\rangle$.

The determinant of the inverse of the Green's function can be simplified as
\begin{align}
	\text{det}(G^{-1}) & =\left|\begin{array}{cc}
		\tilde{\varepsilon}_{f}+a\tau_{z}-\omega & V\\
		V & \tilde{H}_{c}+\varepsilon_{s}-\omega
	\end{array}\right|\nonumber \\
	& =|(\tilde{\varepsilon}_{f}+a\tau_{z}-\omega)(H_{c}+\varepsilon_{s}-\omega)-V^{2}| \nonumber \\
	& =|S^{\dagger}[(\tilde{\varepsilon}_{f}+a\tau_{z}-\omega)(H_{c}+\varepsilon_{s}-\omega)-V^{2}]S|\nonumber \\
	& =|(\tilde{\varepsilon}_{f}+a\tau_{z}-\omega)(\tilde{H}_{c}+\varepsilon_{s}-\omega)-V^{2}|\label{eq:determ}  \\
	& =|(\tilde{\varepsilon}_{f}+a\tau_{z}-\omega)(\bm{u}\cdot\bm{\tau}+\lambda D\sigma_{z}\tau_{z}+\varepsilon_{s}-\omega)-V^{2}|\nonumber \\
	& =\left|\begin{array}{cc}
		b_{0+}+\bm{b}_{+}\cdot\bm{\tau} & 0\\
		0 & b_{0-}+\bm{b}_{-}\cdot\bm{\tau}
	\end{array}\right|\nonumber \\
	& =(b_{0+}^{2}-b_{+}^{2})(b_{0-}^{2}-b_{-}^{2})\nonumber ,
\end{align}
where
\begin{align*}
	b_{0\alpha} & =\omega^{2}-V^{2}-\omega(\tilde{\varepsilon}_{f}+\varepsilon_{s})+\tilde{\varepsilon}_{f}\varepsilon_{s}+au_{z}^{\prime}\\
	b_{x} & =(-\omega+\tilde{\varepsilon}_{f})u_{x}-iau_{y}\\
	b_{y} & =(-\omega+\tilde{\varepsilon}_{f})u_{y}+iau_{x}\\
	b_{z\alpha} & =-\omega(u_{z}+a+\alpha\lambda D)+\tilde{\varepsilon}_{f}u_{z}^{\prime}+a\varepsilon_{s}\\
	b_{\alpha} & =\sqrt{b_{x}^{2}+b_{y}^{2}+b_{z\alpha}^{2}}
\end{align*}
with $u_{z}^{\prime}=u_{z}+\alpha\lambda D$ and $\alpha=\pm1$. In the derivation, we
have used the identity
\begin{equation}
	\det\left(\begin{array}{cc}
		A & B\\
		C & D
	\end{array}\right)=\det(AD-ACA^{-1}B),
\end{equation}
where $A$, $B$, $C$ and $D$ are $n\times n$, $n\times m$, $m\times n$
and $m\times m$ matrices, respectively, and $A$ is invertible. It
follows immediately from the identity
\begin{equation}
	\left(\begin{array}{cc}
		A & B\\
		C & D
	\end{array}\right)\left(\begin{array}{cc}
		I & -A^{-1}B\\
		0 & I
	\end{array}\right)=\left(\begin{array}{cc}
		A & 0\\
		C & D-CA^{-1}B
	\end{array}\right).
\end{equation}
We also have
\begin{equation}
	b_{x}^{2}+b_{y}^{2} =[(-\omega+\tilde{\varepsilon}_{f})^{2}-a^{2}](u_{x}^{2}+u_{y}^{2}).
\end{equation}
The poles of the Green's function are determined by $\text{det}(G^{-1})=0$ which yields
\begin{align*}
	b_{0\alpha}^{2} & =b_{\alpha}^{2}.
\end{align*}
To determine the position of a Weyl point in momentum space, we suppose that $\tilde{\varepsilon}_{f}$
and $a$ are real ($\tilde{\varepsilon}_{f}=\varepsilon_{fr}$ and
$a=a_{1}$). The existence of a Weyl point at zero energy $\omega=0$ requires
that $b_{0}(\omega=0)=b(\omega=0)$ and $b_{0}(\omega=0)=-b(\omega=0)$ so that
\begin{align}
	b_{0}(\omega=0) & =0\\
	b(\omega=0) & =0,
\end{align}
where we have dropped the subscript $\alpha$ to simplify notations.
Specifically, we require that
\begin{align}
	-V^{2}+\tilde{\varepsilon}_{f}\varepsilon_{s}+au_{z}^{\prime} & =0\\
	u_{x}=u_{y} & =0\\
	\tilde{\varepsilon}_{f}u_{z}^{\prime}+a\varepsilon_{s} & =0.
\end{align}
These equations indicate that the location of a Weyl point is the same as that in $H_{c}$
with an effective mass $\tilde{m}=m-u_{z}^{\prime}$. In fact, only $k_{0}$
changes to $\tilde{k}_{0}=2\sin^{-1}(|\tilde{m}/4\lambda|)$. These
equations further lead to
\begin{align}
	u_{z}^{\prime}=u_{Wz} & =\frac{-a_{1}V^{2}}{\varepsilon_{fr}^{2}-a_{1}^{2}}\\
	\varepsilon_{s} & =\frac{\varepsilon_{fr}V^{2}}{\varepsilon_{fr}^{2}-a_{1}^{2}}.
\end{align}

We are now interested in deriving the energy dispersion near a Weyl point. By expanding $u_x$, $u_y$ and $u_z^\prime$
around zero, that it,
$u_{x}=d_{x}$, $u_{y}=d_{y}$ and $u_{z}^{\prime}=u_{Wz}+d_{z}$ where
$d_{x}$, $d_{y}$ and $d_{z}$ are the first-order small quantities, we obtain
\begin{align}
	b_{0} & =\omega^{2}-\omega(\varepsilon_{fr}+\varepsilon_{s})+a_{1}d_{z}\\
	b_{x} & =(-\omega+\varepsilon_{fr})d_{x}-ia_{1}d_{y}\\
	b_{y} & =(-\omega+\varepsilon_{fr})d_{y}+ia_{1}d_{x}\\
	b_{z} & =-\omega(d_{z}+a_{1}+u_{Wz})+\varepsilon_{fr}d_{z}.
\end{align}
Based on these expressions, we derive the energy spectrum around zero energy up to the first
order as
\begin{align}
	\omega & \approx\frac{2d_{z}\varepsilon_{fr}u_{Wz}\pm\sqrt{v_{x}^{2}(d_{x}^{2}+d_{y}^{2})+v_{z}^{2}d_{z}^{2}}}{c_{0}},\label{eq:WeylDis}
\end{align}
where $v_{x}^{2}=(a_{r}^{2}-\varepsilon_{fr}^{2})c_{0}$, $v_{z}^{2}=v_{x}^{2}+4\varepsilon_{fr}^{2}u_{Wz}^{2}$
and $c_{0}=(a_{1}+u_{Wz})^{2}-(\varepsilon_{fr}+\varepsilon_{s})^{2}$.
The result clearly shows the linear dispersion for the energy near the Weyl point.

\subsection{B2. Energy spectra in the presence of the complex energy independent self-energy}
In this subsection, we consider the effects of both the real and imaginary parts in the self-energy.
To derive an analytical result, we assume that $\Gamma_{0}$ and $\Gamma_{1}$ are
first-order small quantities and $a_{1}=0$.
With these approximations, we can derive the energy dispersion close to zero energy up to the first order as
\begin{align}
	\omega & \approx-i\Gamma_{0}v_{0}\pm\sqrt{v_{1}^{2}[d_{x}^{2}+d_{y}^{2}+(d_{z}-i\gamma_{0})^{2}]}\label{eq:WeylExc}
\end{align}
with $v_{0}=\varepsilon_{s}/(\varepsilon_{fr}+\varepsilon_{s})$ $v_{1}=\varepsilon_{fr}/(\varepsilon_{fr}+\varepsilon_{s})$
and$\gamma_{0}=\varepsilon_{s}\Gamma_{1}/\varepsilon_{fr}$. With nonzero
$\Gamma_{1}$, it is easy to see that a Weyl point becomes a Weyl exceptional ring determined
by $d_{z}=0$ and $d_{x}^{2}+d_{y}^{2}-\gamma_{0}^{2}=0$.

To analyze the effects of $a_{1}$, we assume that it is a first-order small quantity
(so is $u_{Wz}$). We find that $a_{1}$ does not affect our results up to the first order.
Since $a_{1}$ is involved in $\varepsilon_{s}$,
one may think that some higher-order corrections from $a_1$ are included in $\varepsilon_{s}$.

\subsection{B3. Renormalization due to the energy-dependent parts in the self-energy}
We now study the effects of the energy dependent parts in the self-energy. The self-energy can be expanded in the Taylor series
up to the first order with respect to $\omega$ as
\begin{equation}
	\Sigma^{f}\approx a_{0}-i\Gamma_{0}+(a_{1}-i\Gamma_{1})\tau_{z}\sigma_{0}+a_{0\omega}\omega+a_{1\omega}\omega\tau_{z}\sigma_{0},
\end{equation}
where $a_{0\omega}$ and $a_{1\omega}$ are complex numbers. The
inverse of the Green's function is
\begin{widetext}
\begin{align}
	G^{-1} & =\omega-\left(\begin{array}{cc}
		\tilde{\varepsilon}_{f}+a\tau_{z}+a_{\omega}\omega+a_{z\omega}\omega\tau_{z} & V\\
		V & H_{c}+\varepsilon_{s}
	\end{array}\right)\\
	& =\left(\begin{array}{cc}
		\begin{array}{cc}
			\omega(1-a_{0\omega}-a_{1\omega})-\tilde{\varepsilon}_{f}-a & 0\\
			0 & \omega(1-a_{0\omega}+a_{1\omega})-\tilde{\varepsilon}_{f}+a
		\end{array} & \begin{array}{cc}
			-V & 0\\
			0 & -V
		\end{array}\\
		\begin{array}{cc}
			-V & 0\\
			0 & -V
		\end{array} & \omega-H_{c}-\varepsilon_{s}
	\end{array}\right).
\end{align}
We now evaluate the determinant of the inverse of the Green's function,
\begin{align}
	\text{det}(G^{-1})= & \left|\begin{array}{cc}
		\begin{array}{cc}
			\omega(1-a_{0\omega}-a_{1\omega})-\tilde{\varepsilon}_{f}-a & 0\\
			0 & \omega(1-a_{0\omega}+a_{1\omega})-\tilde{\varepsilon}_{f}+a
		\end{array} & \begin{array}{cc}
			-V & 0\\
			0 & -V
		\end{array}\\
		\begin{array}{cc}
			-V & 0\\
			0 & -V
		\end{array} & \omega-H_{c}-\varepsilon_{s}
	\end{array}\right|\\
	= & \frac{1}{Z_{A}Z_{B}}\left|\begin{array}{cc}
		\begin{array}{cc}
			\omega-Z_{A}\tilde{\varepsilon}_{f}-Z_{A}a & 0\\
			0 & \omega-Z_{B}\tilde{\varepsilon}_{f}+Z_{B}a
		\end{array} & \begin{array}{cc}
			-\sqrt{Z_{A}}V & 0\\
			0 & -\sqrt{Z_{B}}V
		\end{array}\\
		\begin{array}{cc}
			-\sqrt{Z_{A}}V & 0\\
			0 & -\sqrt{Z_{B}}V
		\end{array} & \omega-H_{c}-\varepsilon_{s}
	\end{array}\right|\\
	= & \frac{1}{Z_{A}Z_{B}}\left|\begin{array}{cc}
		\omega-\bar{\varepsilon}_{f}-\bar{a}\tau_{z} & -(V_{1}+V_{2}\tau_{z})\\
		-(V_{1}+V_{2}\tau_{z}) & \omega-H_{c}-\varepsilon_{s}
	\end{array}\right|\\
	= & \frac{1}{Z_{A}Z_{B}}\left|(\omega-\bar{\varepsilon}_{f}-\bar{a}\tau_{z})(\omega-H_{c}-\varepsilon_{s})-(V_{1}+V_{2}\tau_{z})^{2}\right|\\
	= & \frac{1}{Z_{A}Z_{B}}\left|(\omega-\bar{\varepsilon}_{f}-\bar{a}\tau_{z})(\omega-H_{c}-\varepsilon_{s})+V_{0}\tau_{z}-\bar{V}^{2}\right|,
\end{align}
\end{widetext}
where $Z_{A}=1/(1-a_{0\omega}-a_{1\omega})$, $Z_{B}=1/(1-a_{0\omega}+a_{1\omega})$,
$\bar{\varepsilon}_{f}=\bar{\varepsilon}_{fr}-i\bar{\Gamma}_{0}=[(Z_{A}+Z_{B})\tilde{\varepsilon}_{f}+(Z_{A}-Z_{B})a]/2$,
$\bar{a}=\bar{a}_{1}-i\bar{\Gamma}_{1}=[(Z_{A}-Z_{B})\tilde{\varepsilon}_{f}+(Z_{A}+Z_{B})a]/2$,
$V_{1}=(\sqrt{Z_{A}}+\sqrt{Z_{B}})V/2$, $V_{2}=(\sqrt{Z_{A}}-\sqrt{Z_{B}})V/2$,
$\bar{V}=\sqrt{V_{1}^{2}+V_{2}^{2}}$ and $V_{0}=-2V_{1}V_{2}$. The determinant can be further reduced to
\begin{widetext}
\begin{align}
	\text{det}(G^{-1})
	& =\frac{1}{Z_{A}Z_{B}}\left|S^{\dagger}[(\omega-\bar{\varepsilon}_{f}-\bar{a}\tau_{z})(\omega-H_{c}-\varepsilon_{s})+V_{0}\tau_{z}-\bar{V}^{2}]S\right|\\
	& =\frac{1}{Z_{A}Z_{B}}\left|(\omega-\bar{\varepsilon}_{f}-\bar{a}\tau_{z})(\omega-\tilde{H}_{c}-\varepsilon_{s})+V_{0}\tau_{z}-\bar{V}^{2}\right|,
\end{align}
\end{widetext}
which is almost the same as Eq. (\ref{eq:determ}) except a prefactor
$1/(Z_{A}Z_{B})$ and a new term $V_{0}\tau_{z}$, which can be obtained
by replacing $a\varepsilon_{s}$ with $a\varepsilon_{s}+V_{0}$ in
Eq. (\ref{eq:determ}). We now assume that
$a_{0\omega}$ and $a_{1\omega}$ are real. Similar to the preceding case, when $\bar{\varepsilon}_{f}$
and $\bar{a}$ are real, the existence of Weyl points at zero energy
requires $\omega=0$ and
\begin{align}
	\bar{\varepsilon}_{f}\varepsilon_{s}-\bar{V}^{2}+\bar{a}u_{z}^{\prime} & =0\\
	u_{x}=u_{y} & =0\\
	\bar{\varepsilon}_{f}u_{z}^{\prime}+\bar{a}\varepsilon_{s}+V_{0} & =0
\end{align}
which leads to
\begin{align}
	u_{z}^{\prime}=u_{Wz} & =-\frac{\bar{V}^{2}\bar{a}_{1}+V_{0}\bar{\varepsilon}_{fr}}{\bar{\varepsilon}_{fr}^{2}-\bar{a}_{1}^{2}} \\
	\varepsilon_{s} & =\frac{\bar{\varepsilon}_{fr}\bar{V}^{2}+V_{0}\bar{a}_{1}}{\bar{\varepsilon}_{fr}^{2}-\bar{a}_{1}^{2}}.
\end{align}
Around the Weyl point, one can also derive the energy dispersion,
which is given by Eq. (\ref{eq:WeylDis}) with renormalized parameters
and $\varepsilon_{s}$ and $u_{Wz}$ including extra terms. For clarity,
we write down the dispersion explicitly,
\begin{equation}
	\omega\approx\frac{2d_{z}\bar{\varepsilon}_{fr}u_{Wz}\pm\sqrt{\bar{v}_{x}^{2}(d_{x}^{2}+d_{y}^{2})+\bar{v}_{z}^{2}d_{z}^{2}}}{\bar{c}_{0}},
\end{equation}
where $\bar{v}_{x}^{2}=(\bar{a}_{1}^{2}-\bar{\varepsilon}_{fr}^{2})\bar{c}_{0}$,
$\bar{v}_{z}^{2}=\bar{v}_{x}^{2}+4\bar{\varepsilon}_{fr}^{2}u_{Wz}^{2}$
and $\bar{c}_{0}=(\bar{a}_{1}+u_{Wz})^{2}-(\bar{\varepsilon}_{fr}+\varepsilon_{s})^{2}$.

In the presence of the imaginary parts in $\bar{\varepsilon}_{f}$
and $\bar{a}$, if $\bar{\Gamma}_{0}$, $\bar{\Gamma}_{1}$ $\bar{a}_{0\omega}$
and $\bar{a}_{1\omega}$ are first-order small quantities (so is $u_{Wz}$),
the dispersion is also given by Eq. (\ref{eq:WeylExc}) with renormalized
parameters, that is,
\begin{equation}
	\omega\approx-i\bar{\Gamma}_{0}\bar{v}_{0}\pm\sqrt{\bar{v}_{1}^{2}[d_{x}^{2}+d_{y}^{2}+(d_{z}-i\bar{\gamma}_{0})^{2}]}
\end{equation}
with $\bar{v}_{0}=\varepsilon_{s}/(\bar{\varepsilon}_{fr}+\varepsilon_{s})$
$\bar{v}_{1}=\bar{\varepsilon}_{fr}/(\bar{\varepsilon}_{fr}+\varepsilon_{s})$
and$\bar{\gamma}_{0}=\varepsilon_{s}\bar{\Gamma}_{1}/\bar{\varepsilon}_{fr}$.

\section*{Appendix C: The perturbation theory}
\setcounter{equation}{0} 
\setcounter{table}{0}
\renewcommand{\theequation}{C\arabic{equation}} 
\renewcommand{\bibnumfmt}[1]{[#1]} \renewcommand{\citenumfont}[1]{#1}

\begin{figure*}[htbp]
	\includegraphics[width = 0.6\linewidth]{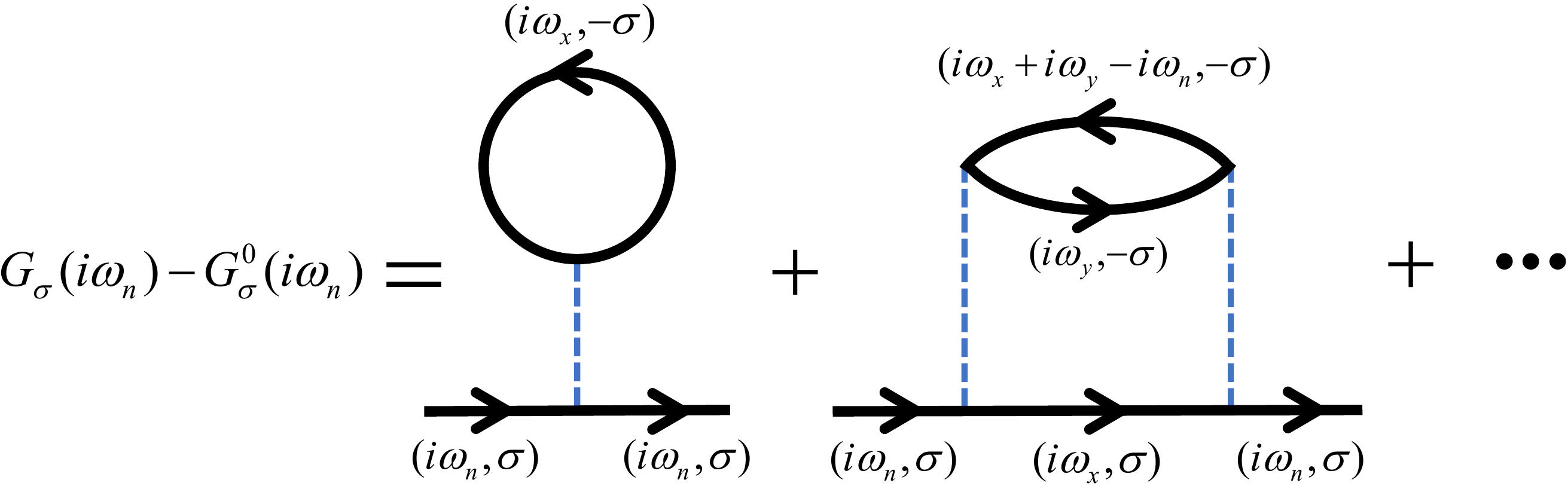}
	\caption{{Diagrammatic expansion for the $f$-electron Matsubara Green's function.}}	
	\label{sigma_green}
\end{figure*}

In this appendix, we compute the self-energy using the second-order perturbation theory. For the interactions in the form of $U\hat{n}_{i,\uparrow}\hat{n}_{i,\downarrow}$, the $f$-electron Matsubara Green's function up to the second-order corrections can be described by the one-particle-irreducible diagram, as shown in Fig.~\ref{sigma_green}. The self-energy up to the second-order corrections is expressed as~\cite{Schweitzer1989,Schweitzer1990}
\begin{widetext}
\begin{align}
	\label{sigmagreen}
	\Sigma_{\sigma,j}(i\omega_n)=&Un^f_{-\sigma}-U^2 T^2\sum_{\omega_x,\omega_y}G^{f}_{\sigma,j}(i\omega_x)G^{f}_{-\sigma,j}(i\omega_y)G^{f}_{-\sigma,j}(i\omega_x+i\omega_y-i\omega_n),
\end{align}
\end{widetext}
where $i\omega_n$ is the Matsubara frequency, $\sigma=\uparrow,\downarrow$ is the spin index, $j=A,B$ is the sublattice index, $T$ is the temperature and $G^{f}_{\sigma,j}(i\omega_n)$ is the corresponding $f$-electron Matsubara Green's function. With time-reversal symmetry,
Matsubara Green's functions of spin up and down have the same form and we thus drop the spin index. The blue dashed line in Fig.~\ref{sigma_green} represents the interaction term connecting four Matsubara Green's functions of $f$ electrons. For any order of perturbation, one can show with the method of the equation of motion~\cite{Gorski2013, Schweitzer1990}, the self-energy must connect with Matsubara Green's functions of $f$ electrons from the same sublattices. It indicates, for the matrix form of self-energy, only diagonal terms of $f$ electrons are nonzero.

\begin{figure*}[htbp]
	\includegraphics[width = 0.7\linewidth]{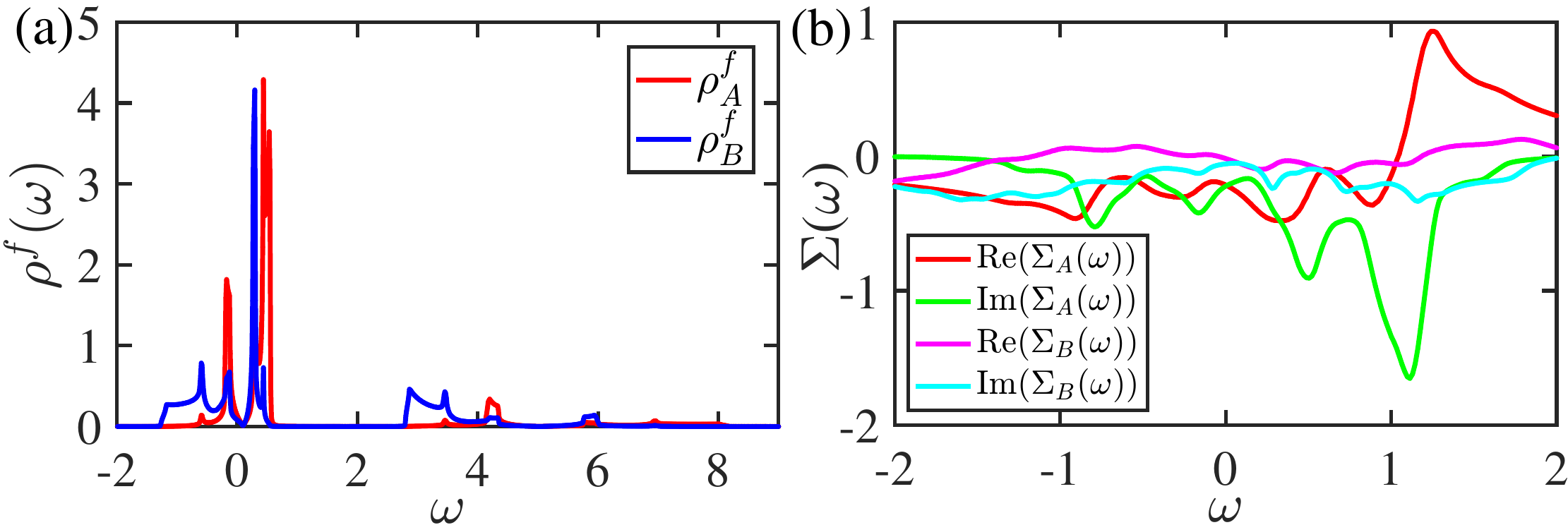}
	\caption{(a) The density of states for the $f$ electrons at sublattices $A$ and $B$
		without interactions as a function of the energy $\omega$.
		(b) The numerically computed self-energy based on Eq.~(\ref{secondsigma}) at the temperature $T=1/6$.
		Here, $t=0.5$, $m=1.2$, $\lambda=0.6$, $V=2$, $U=2$, $\varepsilon_f=1.125$ and $\varepsilon_{s}=4$.}	
	\label{r_a_s}
\end{figure*}
\begin{table*}[h]
	\caption{{List of Taylor coefficients for the self-energy at different temperatures evaluated by Eq.~(\ref{secondsigma}).}
		Here $t=0.5$, $m=1.2$, $\lambda=0.6$, $V=2$, $U=2$, $\varepsilon_f=1.125$, and $\varepsilon_{s}=4$.}
	\centering
	\setlength{\tabcolsep}{3pt}
	\begin{tabular}{c c c c c c c c}
		\toprule
		$T$ & Re($a_{0A}$)&Re($a_{1A}$)&Im($a_{0A}$)&Re($a_{0B}$)&Re($a_{1B}$)&Im($a_{0B}$)&$\Gamma_1$ \tabularnewline
		\midrule
		1/30  &-0.1921 &-0.5401 &-0.0012 &-0.0350 &-0.2381 &$-1.76\times10^{-4}$ & $5.32\times10^{-4}$  \tabularnewline
		\midrule
		1/28  &-0.1932 &-0.5423 &-0.0016 &-0.0353 &-0.2391 &$-2.44\times10^{-4}$ & $6.63\times10^{-4}$  \tabularnewline
		\midrule
		1/26  &-0.1944 &-0.5451 &-0.0020 &-0.0357 &-0.2404 &$-3.48\times10^{-4}$ & $8.19\times10^{-4}$  \tabularnewline
		\midrule
		1/24  &-0.1960 &-0.5486 &-0.0025 &-0.0362 &-0.2422 &$-5.11\times10^{-4}$ & 0.0010  \tabularnewline
		\midrule
		1/22  &-0.1980 &-0.5530 &-0.0033 &-0.0367 &-0.2447 &$-7.65\times10^{-4}$ & 0.0013  \tabularnewline
		\midrule
		1/20  &-0.2004 &-0.5588 &-0.0045 &-0.0372 &-0.2484 &-0.0012 & 0.0016 \tabularnewline
		\midrule
		1/18  &-0.2033 &-0.5662 &-0.0062 &-0.0377 &-0.2537 &-0.0018 & 0.0022 \tabularnewline
		\midrule
		1/16  &-0.2067 &-0.5753 &-0.0092 &-0.0379 &-0.2615 &-0.0030 & 0.0031 \tabularnewline
		\midrule
		1/14  &-0.2107 &-0.5860 &-0.0147 &-0.0377 &-0.2731 &-0.0049 & 0.0049 \tabularnewline
		\midrule
		1/12  &-0.2149 &-0.5963 &-0.0252 &-0.0364 &-0.2899 &-0.0084 & 0.0084 \tabularnewline
		\midrule
		1/10  &-0.2190 &-0.6009 &-0.0457 &-0.0333 &-0.3137 &-0.0148 & 0.0155 \tabularnewline
		\midrule
		1/8   &-0.2222 &-0.5881 &-0.0864 &-0.0274 &-0.3443 &-0.0266 & 0.0299 \tabularnewline
		\midrule
		1/6   &-0.2229 &-0.5415 &-0.1640 &-0.0179 &-0.3756 &-0.0496 & 0.0572 \tabularnewline
		\bottomrule
	\end{tabular}
	\label{Table1}
\end{table*}

The first-order self-energy term can be understood as the Hartree part of the electron's self-energy and is not a function of the frequency
and thus can be contained in $\varepsilon_f$. By performing the sum over the Matsubara frequency, the second-order self-energy can be reduced to~\cite{Schweitzer1989,Schweitzer1990}
\begin{widetext}
\begin{align}
	\label{secondsigma} &\Sigma_{j}(\omega)=-U^2\iiint_{-\infty}^{+\infty}d\omega_1d\omega_2d\omega_3\rho^{f}_{j}(\omega_1)\rho^{f}_{j}(\omega_2)\rho^{f}_{j}(\omega_3)\frac{n_F(\omega_1)n_F(\omega_2)n_F(-\omega_3)+n_F(-\omega_1)n_F(-\omega_2)n_F(\omega_3)}{\omega-\omega_1-\omega_2+\omega_3+i0^+},
\end{align}
\end{widetext}
where $n_F=1/(e^{\omega/T}+1)$ is the Fermi-Dirac distribution function, and $\rho^{f}_{j}(\omega)=-\frac{1}{\pi}\textrm{Im}G^{f}_{j}(\omega+i0^+)$ is $f$-electron density of states at the sublattice $j$
in the absence of interactions.

Figure~\ref{r_a_s}(a) shows the density of states of the $f$-electron $\rho_{A/B}^f(\omega)$ with $t=0.5$, $m=1.2$, $\lambda=0.6$, $V=2$, $U=2$, $\varepsilon_f=1.125$ and $\varepsilon_{s}=-4$, which is used to compute the self-energy. The densities of states
vanish at the energy close to the zero energy, indicating the existence of Weyl points there.
Figure~\ref{r_a_s}(b) displays the numerically calculated second-order self-energy based on Eq. (\ref{secondsigma}) at the temperature $T=1/6$. One can observe that the self-energy exhibits oscillations, which result from the van Hove singularities in the density of states. In addition, the amplitude of the self-energy at the sublattice $A$ is much larger than that at the sublattice $B$
due to the more compact $\rho_A^f(\omega)$.

\begin{figure*}[htbp]
	\includegraphics[width = 1\linewidth]{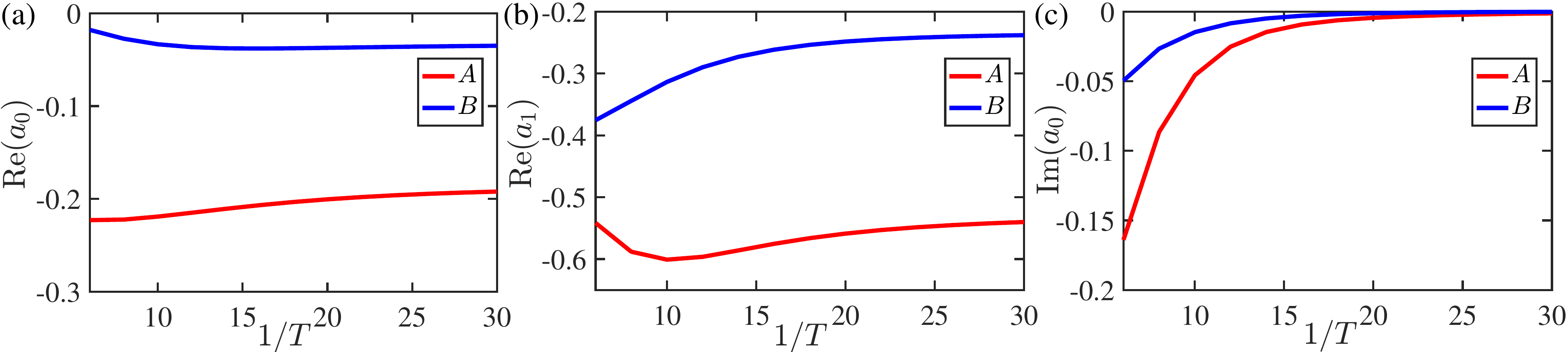}
	\caption{{Plots of Taylor coefficients for the self-energy at different temperatures using the data listed in Table~\ref{Table1}.}
		Here, $t=0.5$, $m=1.2$, $\lambda=0.6$, $V=2$, $U=2$, $\varepsilon_f=1.125$ and $\varepsilon_{s}=4$.}	
	\label{sigma_w_coeff}
\end{figure*}

We calculate the self-energies at different temperatures and perform the Taylor expansion with respect to $\omega$
near the zero energy,
\begin{eqnarray}
	\Sigma^f(\omega) & \approx & a_{0}-i\Gamma_{0}+(a_{1}-i\Gamma_{1})\tau_{z}+a_{0\omega}\omega+a_{1\omega}\omega\tau_{z} \\
	&=& \left(
	\begin{array}{cc}
		a_{0A}+a_{1A}\omega & 0 \\
		0 & a_{0B}+a_{1B}\omega \\
	\end{array}
	\right),
\end{eqnarray}
where $\text{Re}(a_{0A})=a_0+a_1$, $\text{Im}(a_{0A})=-(\Gamma_0+\Gamma_1)$, $\text{Re}(a_{1A})=a_{0\omega}+a_{1\omega}$,
$\text{Re}(a_{0B})=a_0-a_1$, $\text{Im}(a_{0B})=-(\Gamma_0-\Gamma_1)$ and $\text{Re}(a_{1B})=a_{0\omega}-a_{1\omega}$.
The numerically computed Taylor coefficients are listed in Table~\ref{Table1},
where $\text{Im}(a_{1A})$ and $\text{Im}(a_{1B})$ are not displayed as their values are much smaller than those of the corresponding real parts.
To clearly see their change with respect to temperatures, we also provide the curve description in Fig.~\ref{sigma_w_coeff}. As discussed in the preceding section,
the existence of $\text{Re}(a_{0A})$ and $\text{Re}(a_{0B})$ changes the position of Weyl points in momentum space and their energy (if $\varepsilon_s$ is
held fixed). Figure~\ref{sigma_w_coeff}(a) tells us that $\text{Re}(a_{0A})$ and $\text{Re}(a_{0B})$ only slightly change with temperatures,
indicating that the position and energy of Weyl points change slightly.
$\text{Im}(a_{0A})$ [$\text{Im}(a_{0B})$] reveal the inverse of the lifetime of quasiparticles at sublattice $A$ [$B$] and must be negative.
Both $|\text{Im}(a_{0A})|$ and $|\text{Im}(a_{0B})|$ increase with the rise of temperatures, and
their difference $\Gamma_1$ also increases significantly with temperatures, leading to enlarged Weyl exceptional rings as temperatures rise,
which further merge into two exceptional rings as discussed in the main text.
As discussed in the preceding section, $\text{Re}(a_{1A})$ and $\text{Re}(a_{1B})$ renormalize system parameters and thus do not affect the
the qualitative feature of the energy spectrum.

\section*{Appendix D: The spectral functions with respect to the energy} 
\setcounter{equation}{0} 
\setcounter{table}{0}
\renewcommand{\theequation}{D\arabic{equation}} 
\renewcommand{\bibnumfmt}[1]{[#1]} \renewcommand{\citenumfont}[1]{#1}

\begin{figure*}[htbp]
	\includegraphics[width = 0.9\linewidth]{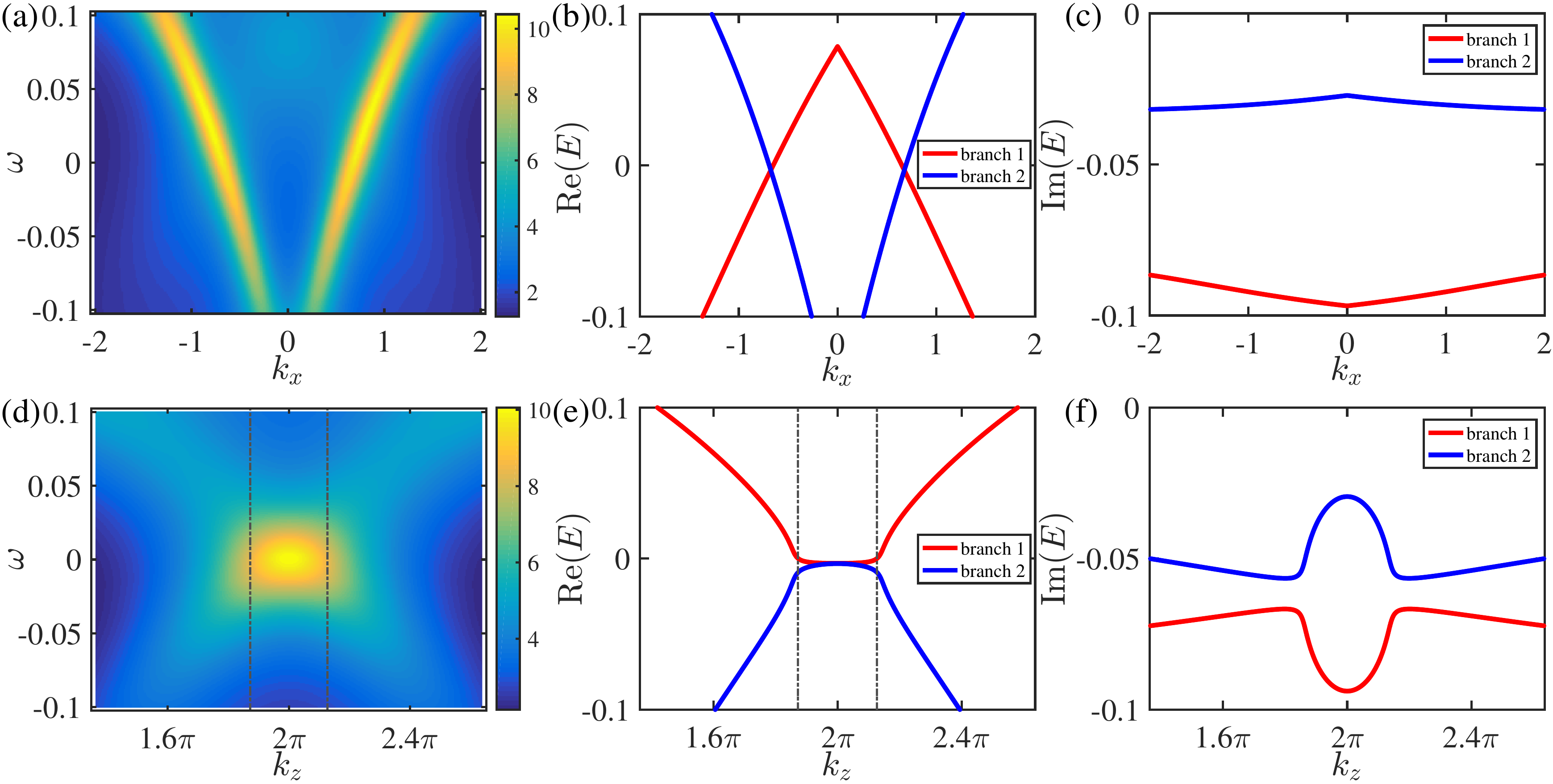}
	\caption{(a) The spectral function $\rho(\omega,\bm{k})$ with respect to $\omega$ and $k_x$ with $k_y=0$ and $k_z=2\pi$.
		The real and imaginary parts of the corresponding energy spectrum for the second and third bands of the effective Hamiltonian are
		plotted in (b) and (c), respectively.
		(d) The spectral function $\rho(\omega,\bm{k})$ with respect to $\omega$ and $k_z$ with $k_x=-0.675$ and $k_y=0$
		with the real and imaginary parts of the corresponding energy spectrum plotted in (e) and (f), respectively.
		In (d) and (e), two vertical lines refer to the positions where the real part of the energy spectrum begins to split.
		Here, $t=0.5$, $m=1.2$, $\lambda=0.6$, $V=2$, $U=2$, $\varepsilon_f=1.125$, $\varepsilon_{s}=4$, and $T=1/6$.}	
	\label{rpv}
\end{figure*}

In the main text, we have shown the spectral functions with respect to the energy at three fixed points in momentum space.
Here, we analyze the features of the spectral functions as functions of both $\omega$ and $\bm{k}$.
We consider two cases: One is along the $k_x$ line with $k_y=0$ and $k_z=2\pi$ which crosses the Fermi tube,
and the other is along the $k_z$ line with $k_x=-0.675$ and $k_y=0$ which is along the Fermi tube.
In Fig.~\ref{rpv}, we plot the spectral functions at the temperature $T=1/6$,
which are numerically calculated by the second-order perturbation theory.
In the former case [see Fig.~\ref{rpv}(a)], there appear two bright lines crossing zero energy corresponding to two exceptional points,
which agree well with the blue branch in the energy spectrum (the poles of the Green's function)
shown in Fig.~\ref{rpv}(b). One may wonder why the other red
branch disappears in the spectral function. To interpret the phenomenon, we plot the imaginary parts of the energy spectrum
in Fig.~\ref{rpv}(c), illustrating that the red branch has larger absolute values of the imaginary parts.
With larger imaginary values, the spectral functions are broader so that this branch is invisible
compared to the blue one with smaller imaginary values. In the latter case, the spectral function exhibits
a bright region around $\omega=0$ which extends along $k_z$ near $k_z=2\pi$,
corresponding to the zero energy part in the energy dispersion [see Fig.~\ref{rpv}(e)].
The energy spectrum then splits into two branches as $k_z$ deviates from the flat region, which can also
be observed in the spectral function. For the splitting parts, the peak becomes wider and weaker since
the corresponding imaginary parts of the energy spectra are larger [see Fig.~\ref{rpv}(f)].
Note that while the positions in the spectral function where the splitting happens are slightly different from those
in the energy spectrum, they are closely related. Also note that the imaginary parts of the two branches do not touch
because the chosen momenta do not cross exceptional rings due to the fact that the Fermi surface slightly deviates a cylinder
shape and takes a shape of a barrel.

\section*{Appendix E: Other Data Analysis about the DMFT calculation} 
\setcounter{equation}{0} 
\setcounter{table}{0}
\renewcommand{\theequation}{E\arabic{equation}} 
\renewcommand{\bibnumfmt}[1]{[#1]} \renewcommand{\citenumfont}[1]{#1}

\begin{figure*}[htbp]
	\includegraphics[width = 1\linewidth]{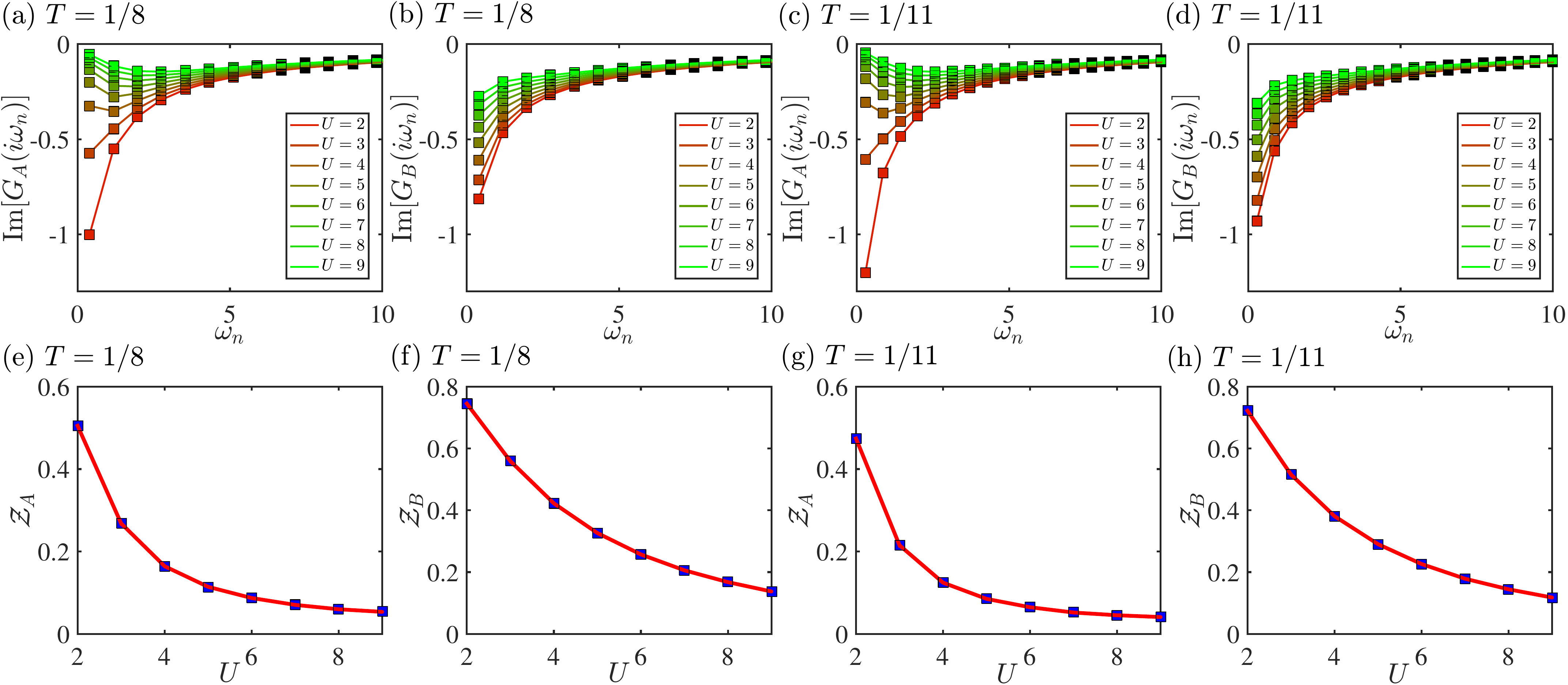}
	\caption{(a-d) Imaginary parts of the Matsubara Green's function and
		(e-h) quasiparticle weights for $f$ electrons on sublattice $A$ or $B$
		with respect to the interaction $U$ at $T=1/8$ or $T=1/11$.}
	\label{Figsm2}
\end{figure*}

To confirm the reliability of our DMFT calculations, we use the existing scripts to compute the Mott transition with the increase of
the interaction strength $U$ at different temperatures. The phase transition can be identified by the imaginary parts of the Matsubara Green's function $G(i\omega_n)$ and the quasiparticle weights $\mathcal{Z}$. A significant decline in the $|\rm{Im}G(i\omega_n)|$ near $\omega_0\equiv \pi T$ is observed in Fig.~\ref{Figsm2}(a-d), which is one of the characteristics when the Mott transition happens. We point out that there is a site-selective Mott-insulating behavior between $A$ and $B$ sites~\cite{Park2012}. While electrons on sublattice $A$ enter into the Mott-insulating phase (e.g., $U>4$), electrons on sublattice $B$ are still in the metallic phase. The distinct behavior arises from the breaking of inversion symmetry,
which
is also crucial for the emergence of different quasiparticle lifetimes on different sublattices.
The Mott transition can also be identified by quasiparticle weights $\mathcal{Z}$, which can be calculated approximately at low temperatures by
\begin{align}
	\label{Z_appro}
	\mathcal{Z}\cong\left[1-\frac{\rm{Im} \Sigma(i\omega_0)}{\omega_0}\right]^{-1}.
\end{align}
The results of $\mathcal{Z}$ are shown in Fig.~\ref{Figsm2}(e-h). We see that with the increase of $U$,
$\mathcal{Z}$ on sublattice $A$ decreases toward zero, signalling a transition from a metallic phase to the Mott-insulating phase.
Compared with the quasiparticle weights on sublattice $A$, the decline of the weights on sublattice $B$ with the interaction is
slower and smoother, which agrees well with the result of the Matsubara Green's function.

\end{document}